\documentclass[runningheads]{llncs}
\usepackage[T1]{fontenc}
\usepackage{graphicx} 
\usepackage{eurosym} 
\usepackage{multirow}
\usepackage[hidelinks]{hyperref}
\usepackage{tabularx}
\usepackage{makecell}
\usepackage{algorithm}
\usepackage{algpseudocode}  
\usepackage{amsmath,amssymb}  
\usepackage{svg}
\usepackage{lscape}    
\usepackage{caption}   
\usepackage{pdflscape}
\usepackage{booktabs}
\usepackage{ragged2e}
\usepackage{comment} 
\usepackage{wrapfig}
\newcommand{\mypar}[1]{\smallskip\noindent\textbf{#1.}}

\begin{document}
\title{Bridging Imperative Process Models and Process Data Queries---Translation and Relaxation}
\titlerunning{Bridging Imperative Process Models and Process Data Queries}
%
\author{Abdur Rehman Anwar Qureshi\inst{1} \and
Adrian Rebmann\inst{2} \and
Timotheus Kampik\inst{2,3}\and
Matthias Weidlich\inst{4}\and
Mathias Weske\inst{1}}
\authorrunning{R. Qureshi et al.}
%
\institute{Hasso Plattner Institute, University of Potsdam, Potsdam, Germany\\ 
\email{\{rehman.qureshi|mathias.weske\}@hpi.de}
\and
SAP Signavio, Berlin, Germany\\
\email{\{adrian.rebmann|timotheus.kampik\}@sap.com}
\and
Umeå University, Umeå, Sweden
\and
Humboldt Universität zu Berlin, Germany\\
\email{matthias.weidlich@hu-berlin.de}}
\maketitle              
\vspace{-1em}
\begin{abstract}
Business process management is increasingly practiced using data-driven approaches. Still, classical imperative process models, which are typically formalized using Petri nets, are not straightforwardly applicable to the relational databases that contain much of the available structured process execution data.
This creates a gap between the traditional world of process modeling and recent developments around data-driven process analysis, ultimately leading to the under-utilization of often readily available process models.
In this paper, we close this gap by providing an approach for translating imperative models into relaxed process data queries, specifically SQL queries executable on relational databases, for conformance checking. 
Our results show the continued relevance of imperative process models to data-driven process management, as well as the importance of behavioral footprints and other declarative approaches for integrating model-based and data-driven process management.

\end{abstract}
\section{Introduction}
\label{introduction}
Imperative process models, such as those specified using the Business Process Model and Notation (BPMN) standard~\cite{omg2011bpmn}, have emerged as the standard approach for modeling and communicating about operational processes. 
These models provide an intuitive abstraction of reality and require moderate training to create. 
However, the shift from model-driven business process management (BPM) to data-driven BPM has prompted growing interest in declarative approaches to process specification~\cite{di2022declarative}. 
This trend is driven by the inherent flexibility of the declarative paradigm, which often better aligns with the complexity and variability of real-world processes. 
Moreover, declarative constraints, which are the central ingredient of these specifications, can be readily translated into queries in well-established languages, such as SQL, that can then be directly executed on process execution data, enabling immediate data-driven insights~\cite{di2022declarative,brandone}.

A key process analysis task that highlights the value of the declarative approach is conformance checking. 
Organizations rely on conformance checking to ensure that operational processes align with desired behavior. 
While imperative models provide a convenient way to represent this behavior, interpreting it in terms of queries allows for more direct and flexible conformance analysis. 
In particular, conformance checking with real-world imperative process models tends to overwhelm the user with many deviations, due to the models' overly rigid nature that does not reflect a potentially more relaxed \emph{intended} meaning.
For instance, while an invoice should ideally be checked after it has been received, real-world processes might involve intermediate steps such as asking for clarification or responding to a vendor inquiry. 
Declarative approaches naturally capture such flexibility, while allowing for immediate and efficient checking of conformance based on queries. This dual need to (1) retain the convenience of imperative modeling while (2) leveraging the flexible querying capabilities of declarative approaches underscores the importance of bridging the two paradigms.

This bridge, however, is not straightforward. 
A clear understanding of how the behavior specified in imperative models can be translated into process data queries is essential but underexplored. 
Although previous studies~\cite{barbaro2025sound,DBLP:conf/bpm/BergmannRK23,goulart2024mining} have translated imperative models into declarative specifications, these efforts lack a demonstration of the relationship to queries and do not consider a systematic relaxation of constraints for conformance checking. 
This limits the ability to seamlessly integrate imperative models and process data querying.

In this paper, we address this gap through a structured approach to bridge imperative process models and queries. 
Specifically, we leverage the trace semantics of sound, free-choice workflow nets, representing them through the well-known directly-follows relations~\cite{DBLP:conf/apn/WeidlichW12}. 
Based on these, a user can systematically relax the allowed behavior by changing relations to accommodate the flexibility of real-world processes thus allowing for deviations that remain logically consistent with the \emph{intended} behavior of the original model.
The remaining relations provide the foundation for deriving a corresponding set of declarative constraints, naturally translate into process data queries (e.g., in plain SQL), which can be directly executed on event data for conformance checking.

The remainder is structured as follows. \autoref{sec:motivation} presents a motivating example to illustrate the practical problem we address. 
\autoref{sec:prelim} introduces the concepts our work is built on. 
\autoref{sec:approach} details our approach, explaining how we determine and relax behavioral relations to generate declarative constraints. 
\autoref{sec:evaluation} provides a proof of concept by applying our approach on real-world data. 
\autoref{sec:related} discusses related research and
 \autoref{sec:conclusion} concludes the paper.

\section{Motivating Example}
\label{sec:motivation}

\begin{figure}[htbp]
    \centering
    \includegraphics[width=\textwidth]{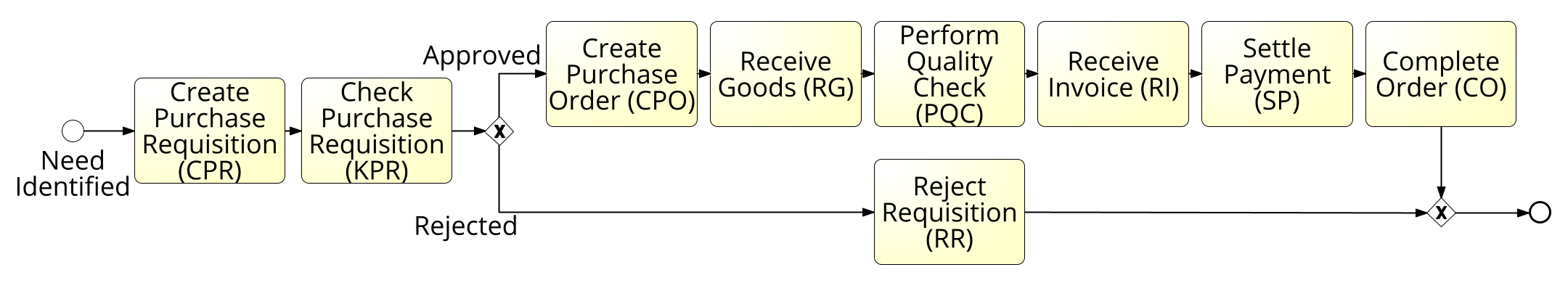}
    \vspace{-1em}
    \caption{An exemplary BPMN process model of a procurement process.}
    \label{fig:example_bpmn}
    \vspace{-1em}
\end{figure}

In this section, we illustrate the value of translating imperative process models into more relaxed process data queries using the process model in \autoref{fig:example_bpmn}.

The process model outlines a procurement workflow, beginning with the creation and approval of a purchase requisition, which triggers the generation of a purchase order. 
Following this, the model requires that goods must be received and then pass a mandatory quality check. Only after this inspection is complete can the invoice be received, leading to the final step of settling the payment.

\begin{table}[h!]
\vspace{-2.5em}
\small
\centering
\caption{One case of the process for a standard office supply purchase.}
\label{tab:event_log}
\begin{tabular}{lll}
\toprule
\textbf{Timestamp}      & \textbf{Activity}              & \textbf{Notes}                             \\
\midrule
2023-11-10 10:00:00   & Create Purchase Requisition  & Order for standard printer paper         \\
2023-11-10 11:30:00   & Approve Purchase Requisition & Auto-approved due to low value           \\
2023-11-10 11:35:00   & Create Purchase Order        &                                          \\
2023-11-13 14:00:00   & Receive Goods                & 1 box of printer paper arrived           \\
2023-11-13 15:10:00   & Receive Invoice              & Invoice received via email               \\
2023-11-20 09:00:00   & Settle Payment               &                                                   \\
\bottomrule
\end{tabular}
\vspace{-1.5em}
\end{table}

In real-world settings, processes rarely adhere perfectly to such rigid specifications. 
There are often exceptions, for instance, to enhance efficiency. 
Consider the event log for a single case in~\autoref{tab:event_log}, which details the purchase of standard office supplies from a trusted vendor. For a low-value, routine item, performing a formal, documented quality check is inefficient and unnecessary. The process owner has thus allowed this step to be skipped to save time and resources. However, when a conformance checking approach aligns the event log in \autoref{tab:event_log} to the BPMN model in \autoref{fig:example_bpmn}, it will detect deviations, namely skipping the purchase requisition and quality checks and the order completion. 

A possible solution is to shift from an imperative to a declarative paradigm. Instead of updating the rigid model for every exceptional situation that may occur, a declarative model that specifies a more relaxed set of constraints is used.
For instance, from the model the following relaxed rule set can be derived: (1) \emph{Create Purchase Requisition} must occur, (2) \emph{Check Purchase Requisition} must precede \emph{Approve Purchase Requisition}, (3) \emph{Receive Invoice} must follow \emph{Receive Goods} (4) \emph{Perform Quality Check} is optional, however, if it occurs, it must occur after \emph{Receive Goods} and before \emph{Settle Payment} (5) \emph{Settle Payment} must occur after \emph{Receive Invoice}.
Checking conformance of the case in \autoref{tab:event_log} against these rules, none of them are violated. In particular, also Rule 4 is not violated because its precondition, the occurrence of \emph{Perform Quality Check}, is not met.

Imperative models, while crucial for designing processes, are often too rigid for real-world conformance checking, flagging any exception as a deviation. The declarative paradigm focuses on core constraints over predefined paths, offering more flexible conformance checking.

\section{Preliminaries}
\label{sec:prelim}
\mypar{Sound, Free-Choice Workflow Net}
A workflow net (WF-net) is a specific type of Petri net. Formally, it is a tuple $W = (P, T, F)$, where P is a set of places, T is a set of transitions, and F is the flow relation between them. To qualify as a WF-net, the structure must satisfy three key properties (1) \emph{A Single Start Point:} The net must contain one unique \emph{source} place $i$ that has no incoming transitions ($\bullet i = \emptyset$). (2) \emph{A Single End Point:} It must also contain one unique \emph{sink} place $o$ that has no outgoing transitions ($o\bullet = \emptyset$). (3) \emph{Full Connectedness:} Every other node (place or transition) must lie on a path from the start place $i$ to the end place $o$. A \emph{sound, free-choice workflow net} adds two crucial guarantees. \emph{Soundness} ensures that the process can always complete properly, without deadlocks or leftover tasks. \emph{Free-choice} ensures that when multiple transitions compete for tokens from the same place, that choice is self-contained and not dependent on other parts of the workflow net.

\mypar{Behavioral Relations}
Alpha-relations offer a fundamental abstraction of process behavior, applicable to both event logs and process models, and have been a cornerstone of process mining since its inception. When inferred from sound, free-choice workflow nets, they describe behavioral relationships between two activities $a, b \in A$ as follows:
\textit{Directly-follows ($a \rightarrow b$):} Activity $a$ can be immediately followed by activity $b$. \textit{Concurrency ($a \parallel b$):} Activities $a$ and $b$ can occur in any order. \textit{Choice ($a - b$):} Activities $a$ and $b$ never follow each other directly.

The transitive closure of the directly-follows relation ($ \rightarrow $) links directly-follows ($ \rightarrow $) and eventually-follows ($ \prec $) relations. This means that $p \prec q$ if there exists a path of one or more \emph{directly-follows} arrows leading from $p$ to $q$. For instance, if $p \rightarrow a$, $a \rightarrow b$, and $b \rightarrow q$, then $p \prec q$. We write $\texttt{TC}(D)$ for the function that computes the transitive closure of directly-follows relations in $D$, which means it identifies all pairs of activities $(p, q)$ where $q$ can be eventually reached from $p$ through a sequence of directly-following activities.

\mypar{From Workflow Net to Directly-Follows Relations}
For sound, free-choice workflow nets, it has been shown that the directly-follows (there called \emph{1-successor}) relations can be inferred directly from the net's structure~\cite{DBLP:conf/apn/WeidlichW12}. 
The approach is based on the concept of the Minimal Structural Successor (MSS) function, which for any two transitions $x$ and $y$, identifies the minimal set of nodes structurally required to enable $y$ after $x$ has fired. 
As proven in their work, $y$ is a 1-successor of $x$ if and only if the corresponding MSS set contains exactly one transition, namely the source transition $x$ itself, formally $|mss(x, y) \cap T| = 1$. 
This guarantees that no other transition is required in between $x$ and $y$. Therefore, by calculating the MSS for all transition pairs, the directly-follows relations that define the same trace semantics as the net can be determined without exhaustive trace generation or state-space exploration.

\mypar{Declarative Constraints}
We translate behavioral relations into declarative constraints focusing on (\emph{Branched})~\emph{Declare}~\cite{pesic2008}, which is a language that abstracts from  \emph{Linear Temporal Logic on Finite Traces ($LTL_f$)}~\cite{de2013linear}. 
It allows us to specify constraints that must hold over the sequence of activities and is built on several core temporal operators, where $\phi$ represents a logical condition \textbf{G}$\phi$ (Globally): $\phi$ must hold at every step of the trace. \textbf{X}$\phi$ (Next): $\phi$ must hold in the immediately following step. $\phi_1$\textbf{U}$\phi_2$ (Until): $\phi_1$ must hold until $\phi_2$ eventually holds. \textbf{F}$\phi$ means that $\phi$ holds at
some point in the future. Table~\ref{tab:semantics-constraints} summarizes the semantics of selected \emph{Declare} templates, whose parameters are sets. This means that any one of those activities can activate resp. satisfy the constraint. For example, the template \texttt{AlternateResponse}($\{a, b\}, \{c\}$) defines that if activity $a$ or ($\lor$) activity $b$ occurs, c must happen afterwards. This provides more flexibility than standard \emph{Declare}.

\begin{table}
\vspace{-1.5em}
\caption{Semantics of the declarative constraints}
\label{tab:semantics-constraints}
\begin{tabular}{p{2.9cm}|p{4cm}|p{5cm}} 
\hline
\textbf{Constraint} & \textbf{LTLf formula} & \textbf{Description} \\
\hline
\texttt{Init}($\{p_1,...,p_n\}$) &  \textbf{G}(\textbf{start}$ \rightarrow$\textbf{X}($\{p_1 \vee p_2  \vee p_n\}$))& Assert that at least one activity from $\{p_1,p_2,...p_n\}$ occurs in start.\\
\hline

\texttt{ChainResponse} ($\{p_1,...p_n\}$, $\{q_1,...,q_n\}$) & $\textbf{G}(\{p_1 \vee \dots \vee p_n\} \rightarrow \textbf{X}(\{q_1 \vee \dots \vee q_n\}))$& Whenever at least one activity from $\{p_1,...,p_n\}$ occurs, then one of $\{q_1,...,q_n\}$ must occur in the immediately following step.\\
\hline

\texttt{AlternateResponse} ($\{p_1,...,p_n\}$, $\{q_1,...,q_n\}$) & \textbf{G}($\{p_1 \vee \dots \vee p_n\} \rightarrow$\textbf{X}($\neg\{p_1 \vee \dots \vee p_n\}$\textbf{U} $\{q_1,\vee \dots \vee,q_n\}$))& Whenever at least one activity from $\{p_1,...,p_n\}$ occurs, then one of $\{q_1,...,q_n\}$ occurs in afterwards, and not a single activity from $\{p_1,...,p_n\}$ recurs in between.\\

\hline
\end{tabular}
\vspace{-1.5em}
\end{table}

\section{Approach}
\label{sec:approach}

In this section, we give an overview of our approach, before detailing its steps.

\begin{figure}
    \centering
    \includegraphics[width=.85\linewidth]{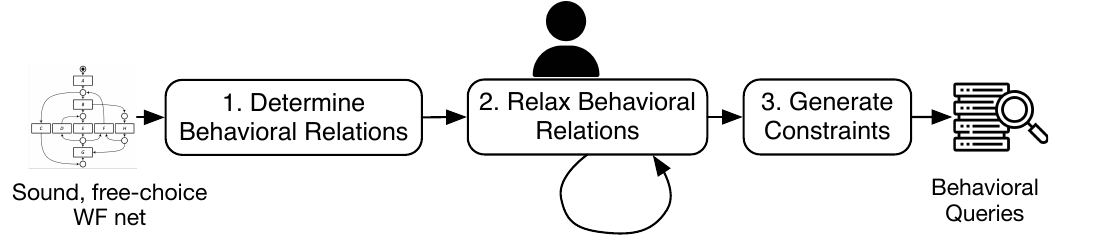}
    \vspace{-1.5em}
    \caption{Approach overview.}
    \label{fig:approach overview}
    \vspace{-1.5em}
\end{figure}

\subsection{Approach Overview}
We present a high-level overview of our approach in \autoref{fig:approach overview}. 
It takes as input a sound, free-choice WF-net (as defined in \autoref{sec:prelim}), which can be obtained from a BPMN diagram using the established method by Dijkman and Ouyang~\cite{dijkman2008semantics}. The approach then consists of three steps: 
(1) determining behavioral relations, 
(2) relaxing behavioral relations (done by a user), and 
(3) generating declarative constraints. 
The output is a set of behavioral process data queries that can be straightforwardly created from the declarative constraints.

\subsection{Step 1: Determine Behavioral Relations}
Step 1 begins by determining directly-follows relations from a sound, free-choice workflow net provided as input. To this end, it employs the method presented in~\cite{DBLP:conf/apn/WeidlichW12} (detailed in Section~\ref{sec:prelim}). Subsequently, based on these directly-follows relations, Step 1 derives Alpha-relations. The output of Step 1 is a relation matrix, as shown in \autoref{tab:relations} for the workflow net from our running example. Crucially, these resulting relations preserve the same trace semantics as the input net.

\begin{table}[htbp]
\vspace{-2em}
    \centering
    \small
    \caption{Behavioral relation matrix of the running example.}
    \label{tab:relations}
    \begin{tabular}{|c|*{9}{c|}}
        \hline
        & \textbf{CPR} & \textbf{KPR} & \textbf{CPO} & \hphantom{I}\textbf{RG\hphantom{I}} & \textbf{PQC} & \hphantom{I}\textbf{RI}\hphantom{I} & \hphantom{I}\textbf{SP}\hphantom{I} & \hphantom{I}\textbf{CO}\hphantom{I} & \hphantom{I}\textbf{RR}\hphantom{I} \\
        \hline
        \textbf{CPR} & - & $\rightarrow$ & - & - & - & - & - & - & - \\
        \hline
        \textbf{KPR} & $\leftarrow$ & - & $\rightarrow$ & - & - & - & - & - & $\rightarrow$ \\
        \hline
        \textbf{CPO} & - & $\leftarrow$ & - & $\rightarrow$ & - & - & - & - & - \\
        \hline
        \textbf{RG} & - & - & $\leftarrow$ & - & $\rightarrow$ & - & - & - & - \\
        \hline
        \textbf{PQC} & - & - & - & $\leftarrow$ & - & $\rightarrow$ & - & - & - \\
        \hline
        \textbf{RI} & - & - & - & - & $\leftarrow$ & - & $\rightarrow$ & - & - \\
        \hline
        \textbf{SP} & - & - & - & - & - & $\leftarrow$ & - & $\rightarrow$ & - \\
        \hline
        \textbf{CO} & - & - & - & - & - & - & $\leftarrow$ & - & - \\
        \hline
        \textbf{RR} & - & $\leftarrow$ & - & - & - & - & - & - & - \\
        \hline
    \end{tabular}
    \vspace{-2em}
\end{table}

\subsection{Step 2: Relax Behavioral Relations}
Next, we introduce types of operations that a process analyst can apply to perform relaxations of the initial behavioral relations of the input model. Once a relaxation operation is applied, our approach updates the matrix accordingly. In this manner, it is transparent how other behavioral relations are affected by the relaxation, which allows a user to judge if this is reasonable for their process analysis and make adjustments if needed.

\mypar{(1) Remove Activity}
This operation makes an activity optional, while also allowing that it can appear whenever. This enables a user to accommodate context-dependent steps. As seen in the running example, skipping the \emph{Perform Quality Check} for a routine purchase is a valid efficiency gain. This prevents such cases from being flagged as deviations.
This is achieved by setting all its relations with other activities to \emph{undirected, eventually-follows} (denoted as $\prec\succ$).

\mypar{(2) Remove All Relationships Between Two Activities}
This operation removes all ordering relations between two activities, making them completely independent. 
This is useful when process fragments are less coupled than the original model suggests. For instance, if a model enforces \emph{Receive Goods $\rightarrow$ Receive Invoice}, this relaxation allows to accommodate real-world scenarios where the invoice arrives before the goods, e.g., when the two packages are sent by different departments.
Their relationship is set to $\prec\succ$ thereby allowing (but nor forcing) them to occur in any order both directly and indirectly.

\mypar{(3) Turn Exclusive (-) into Direct Relationship ($\rightarrow$)}
This operation replaces a choice between activities (\emph{A - B}) with a strict sequence (\emph{A $\rightarrow$ B}). This covers scenarios where activities, once considered exclusive, can co-exist. For instance, if a process allows either an \emph{Automated Approval} or a \emph{Manual Approval}, this relaxation would handle cases where \textit{both} occur for auditing purposes, preventing the system from flagging such valid behavior as a deviation. Note that this also allows for an activity to be followed by itself (e.g., introducing \emph{A $||$ A}), a relaxation essential for modeling rework or iterative tasks. For instance, if for a \emph{Purchase Order} several \emph{Invoices} are created, a rigid model would flag the resulting self-loop as a deviation. This relaxation considers such behavior as an acceptable part of the process, thereby reducing conformance noise.

\mypar{(4) Turn Direct ($\rightarrow$) into Indirect Relationship ($\prec$)}
This operation allows that \emph{B} can also indirectly follow \emph{A}. 
This is critical for conformance, as real-world processes often contain unmodeled steps. For example, an employee might \emph{Log Goods in System} between \emph{Receive Goods} and \emph{Perform Quality Check}. By removing the direct relationship, the behavior gains the flexibility to accept such valid traces.

\subsection{Step 3: Generating Declarative Constraints}
After the user applied a series of relaxation operations, the approach automatically generates a set of declarative constraints as follows.

First, our approach creates a set of directly-follows relations $D$ and a set of eventually-follows relations $E$
based on the (relaxed) matrix. All relations are represented as pairs $(a,b)$ of activities, where bidirectional relations become two individual relations ($(a,b)$ and $(b,a)$).
Then, it generates constraints as outlined in Algorithm \autoref{alg:declarative_constraints}: First, the algorithm initializes the constraint set with one \texttt{Init} constraint using the \texttt{getStartActivities} function. It determines the set of start activities, $I$, by identifying all activities $a \in A$ that are never a successor, i.e., $I = \{ a \in A \mid \forall b \in A, (b, a) \notin D \}$. 
Then the algorithm continues with generating binary constraints. 
Its default template is \texttt{ChainResponse}, as it models the directly-follows relation by enforcing that if an activity \textit{a} occurs, it must be immediately followed by one activity of a given set of successors (lines 4 and 6). 
However, this constraint's inherent OR-logic is too permissive to model parallel behavior, where multiple activities must all occur following a predecessor. For these cases, the algorithm  switches to generating separate \texttt{AlternateResponse} constraints for each parallel activity, which correctly enforces that the predecessor's occurrence requires the eventual occurrence of each successor (lines 7--16). 
This step is further refined to handle optional tasks. To avoid wrongly enforcing a parallel activity that can be skipped, the algorithm employs the \textsc{IsOptionalActivity} function (lines 8 and 11). Before creating an \texttt{AlternateResponse} constraint, it checks if the target activity can be bypassed and---if so--- does not create the constraint, correctly preserving the activity's optionality.

The Boolean function \textsc{IsOptionalActivity} takes an activity $x$ and the directly-follows relations $D$ as input. 
It returns \emph{true} if (and only if) $x$ is identified as optional (skipable) based on a bypass pattern: for every direct predecessor $a$ of $x$ and every successor $b$ of $x$ that are not in parallel relation with $x$ and each other, i.e., $(x,a) \notin D \land (b,x) \notin D \land (b,a) \notin D$, it checks if any $(a,b) \in D$.

Finally, the algorithm processes the eventually-follows relations in the set $E$ to generate \texttt{AlternateResponse} constraints if these are not already covered by the transitive closure of the directly-follows relations (lines 17--21).

\begin{algorithm}
\caption{Generate constraints}
\label{alg:declarative_constraints}
\begin{algorithmic}[1]
\Require $D, E \subseteq A \times A$,  sets of directly-follows and eventually-follows relations (respectively)
\Ensure $C$, a set of declarative constraints

\State $C \gets \{\texttt{Init}(\Call{getStartActivities}{D})$\}
\State $A_D \gets \{\, a \mid (a, b) \in D \,\}$ \Comment{Get unique source activities for direct-follows}

\For{each $a \in A_D$}
    \State $S \gets \{\, x \mid (a, x) \in D \,\}$ \Comment{Collect all directly-following activities}
    \State $C \gets C \cup \{ \texttt{ChainResponse}(\{a\}, S) \}$ \Comment{Generate \texttt{ChainResponse} for $a$}
    \For{each $\{b,c\} \subseteq S$}
        \If{$(b,c) \in D \land (c,b)\in D$} \Comment{Generate dedicated \texttt{AlternateResponse} constraints for parallel relations}
            \If{$\neg \Call{IsOptionalActivity}{b, D}$}
                \State $C \gets C \cup \{ \texttt{AlternateResponse}(\{a\}, \{b\})\}$
            \EndIf
            \If{$\neg \Call{IsOptionalActivity}{c, D}$}
                \State $C \gets C \cup \{ \texttt{AlternateResponse}(\{a\}, \{c\})\}$
            \EndIf
        \EndIf
    \EndFor
\EndFor

\State $A_E \gets \{\, a \mid (a, b) \in E \,\}$ \Comment{Get unique source activities for eventually-follows}
\For{each $a \in A_E$}
    \State $S \gets \{\, x \mid (a, x) \in E \land (a,x) \notin \texttt{TC}(D) \,\}$ \Comment{Collect successors from $E$ that are not in the transitive closure of $D$}
    \State $C \gets C \cup \{ \texttt{AlternateResponse}(\{a\}, S) \}$ \Comment{Generate \texttt{AlternateResponse} for $a$}
\EndFor

\State \Return $C$
\end{algorithmic}
\end{algorithm}

\begin{table}[htb]
\vspace{-1em}
\centering
\caption{Selection of declarative constraints and corresponding SQL fragments. All queries use \texttt{SELECT case\_id FROM events MATCH\_RECOGNIZE}, \texttt{PARTITION BY case\_id ORDER BY end\_time}, and \texttt{ONE ROW PER MATCH}.}
\label{tab:declarative}
\begin{tabularx}{\columnwidth}{@{}l X@{}}
\toprule
\textbf{Constraint} & \textbf{SQL Fragment (\texttt{PATTERN} and \texttt{DEFINE})} \\
\midrule

\texttt{Init(\{${\text{CPR}}$\})} & 
\small\texttt{PATTERN (\^{}CPR ANY*)} \newline
\small\texttt{DEFINE CPR AS event\_name = 'CPR'} \\
\addlinespace 

\texttt{ChainResponse(\text{\{KPR\}, \{CPO,RR\}}}) & 
\small\texttt{PATTERN (ANY* KPR (CPO | RR) ANY*)} \newline
\small\texttt{DEFINE KPR AS event\_name = 'KPR', CPO AS event\_name = 'CPO', RR AS event\_name = 'RR'} \\
\addlinespace

\texttt{ChainResponse({\text{\{RI\}, \{SP\}}}}) & 
\small\texttt{PATTERN (ANY* RI SP ANY*)} \newline
\small\texttt{DEFINE RI AS event\_name = 'RI', SP AS event\_name = 'SP'} \\

\bottomrule
\end{tabularx}
\end{table}

Selected declarative constraints and corresponding SQL queries for the running example (no relaxation) are presented in \autoref{tab:declarative}. The SQL queries can straightforwardly be created from the \emph{Declare} constraints (cf.~\cite{brandone} for details) assuming the following database schema: an event log is represented in a single \emph{events} table, where each row represents a distinct event. 
The table has three columns: the \texttt{case\_id} serves to group events into process instances, the \texttt{end\_time} timestamp orders the events, and the \texttt{event\_name} provides the activity label.

\section{Proof of Concept}
\label{sec:evaluation}

This section demonstrates the practical value of our approach in a conformance checking scenario. We apply our approach to a model of a real-world process and its corresponding public event log. 
To validate the benefits of our novel relaxation step, we compare the resulting conformance insights against traditional alignment-based conformance checking.

\mypar{Implementation}
We implemented the proposed approach as a Python-based command-line interface. 
The source code is publicly available.\footnote{\url{https://github.com/rehman-qureshi/bridging-models-and-queries}}

\mypar{Data}
For our proof of concept, we use the BPI Challenge 2019 (BPIC19) dataset\footnote{\url{https://icpmconference.org/2019/icpm-2019/contests-challenges/bpi-challenge-2019}}. We selected the \emph{3-way matching, invoice before goods receipt} model from the winning submission by Diba et al.~\cite{diba2019compliance}; it covers 86.36\% (217,406) of the 251,734 cases in the original event log. 
To prepare this model for our analysis, we first converted it into a sound, free-choice workflow net—the required input format for our approach—using the method by Dijkman and Ouyang~\cite{dijkman2008semantics}.

\mypar{Setup}
We check conformance using queries that we obtain when applying our approach with relaxation to the BPIC19 workflow net and compare the results with 
\emph{Alignments}, a state-of-the-art conformance-checking technique \cite{van2012replaying}, which aligns event log traces to an imperative process model, i.e., the same net from which our approach generates queries.

\mypar{Approach Application}
We apply our approach to generate behavioral relations, relax these, and generate declarative constraints that map to SQL queries. Based on the initial behavioral relation matrix (after Step 1), we apply the following relaxations for the purpose of this proof-of-concept:

\begin{itemize}
    \item \textit{Recurring Invoicing and Deliveries:} To reflect valid business patterns where multiple invoices may precede goods receipt, or multiple deliveries occur before a single invoice, we first relax ``Record Invoice Receipt'' and ``Record Goods Receipt'' by turning direct relation (\texttt{$||$}) into indirect one (\texttt{$\prec\succ$}), via relaxation type (4), which allows for these to reoccur or be interleaved.
    \item \textit{Decoupling Early Procurement Stages:} We then loosen constraints in the initial stages of the procurement process to allow for greater independence. Specifically, we remove the relationship (relaxation type (2)) between ``Create Purchase Order Item'' and ``Record Order Confirmation'' to allow for independent order confirmation. Furthermore, to permit flexible adjustments of price and quantity before goods movement and invoice handling, we remove all relations between ``Change Price'', ``Change Quantity'', and related purchase requisition/order activities (``Create Purchase Requisition Item'', ``Create Purchase Order Item'', ``Record Order Confirmation''). 
\end{itemize}

These relaxations result in behavioral relations that, 
after translated into constraints and finally queries, considerably reduce the total number of violations by allowing for behavior that is reasonable, allowing analysts to focus on violations of rules that actually matter to the analyst.

\mypar{Results}
\autoref{tab:violations} reports the conformance rate (\# of traces conforming to the model resp.\ all constraints divided by total \# of traces) for both approaches.

\begin{table}[h]
    \vspace{-1em}
  \centering
  \small
  \caption{Conformance rates.}
  \label{tab:violations}
  \begin{tabular}{lc} 
    \hline
    \textbf{Approach} & \textbf{Conformance Rate} \\
    \hline
    Alignments &  0.691 \\
    \hline
    Relaxed queries &  0.823\\
    \hline
  \end{tabular}
  \vspace{-1em}
\end{table}

The numbers illustrate the substantial decrease of detected deviating traces after relaxation, reducing the cognitive load of users that need to investigate which of these are indeed problematic. 

\section{Related Work}
\label{sec:related}
Our work primarily relates to the translation of imperative process models into declarative specifications, relaxation in conformance checking, declarative conformance checking, and the direct querying of process data.

\mypar{From Imperative to Declarative Models} 
Previous work has focused on extracting declarative constraints from BPMN diagrams~\cite{DBLP:conf/bpm/BergmannRK23} and Petri nets~\cite{goulart2024mining}. Barbaro et al.\ present the first approach to systematically turn safe and sound Workflow nets into declarative specifications~\cite{barbaro2025sound}. While their goal is to achieve equivalence of the trace semantics of both models, the goal of our approach is to provide a means for systematically relaxing the allowed behavior instead.

\mypar{Relaxation in Conformance Checking}
A certain degree of relaxation can also be realized within alignment-based conformance checking by manipulating the cost function~\cite{carmona2018conformance}. 
By setting the cost of a log move for an activity to zero, that activity is ignored (optional). While this can implement some simple relaxations, its capabilities are limited. For instance, if two formerly exclusive activities should now be allowed to co-occur, we would require a context-sensitive cost function. This would mean the cost of a log move for one activity would be zero only if the other activity had already occurred in the trace. Such dynamic, context-aware costing is not achievable with a traditional, static cost function.

\mypar{Declarative Conformance Checking}
The use of declarative languages like \emph{Declare}~\cite{pesic2007declare} for conformance checking has gained significant traction for its ability to handle process flexibility. Methods range from providing a simple boolean (conforming/non-conforming) result \cite{chiariello2022asp} to reporting detailed statistics, such as activations, satisfactions, and violations, for each constraint and trace \cite{burattin2016conformance,montali2014monitoring,maggi2019compliance}.

\mypar{Process Querying using SQL}
The direct analysis of event data stored in relational databases is a growing area of interest. Generating SQL queries as done in our approach, builds on applications of native SQL features---specifically, the \texttt{MATCH\_RECOGNIZE} clause---to pattern matching in event logs~\cite{brandone}.

\medskip
Overall, prior work has not systematically bridged imperative models with process data querying. Our approach fills this gap by translating imperative models into executable queries with user-driven relation relaxation, enabling more flexible and practical conformance checking.

\section{Conclusion}
\label{sec:conclusion}

We presented a structured approach to translate sound, free-choice workflow nets into declarative constraints and process data queries based on their underlying behavioral relations, bridging imperative models and data-driven analysis. By allowing users to systematically relax the relations, our approach accommodates real-world process flexibility and reduces conformance checking noise. Our proof-of-concept shows a considerable reduction in violations, which enables analysts to focus on genuine non-conformance. This work highlights the continued relevance of imperative models as a source of process knowledge while leveraging the flexibility of declarative models for scalable data-driven analysis.

While the main purpose of transforming the behavioral matrix into declarative constraints is to ease subsequent query relaxation, a notable limitation is that it remains to be demonstrated whether or not the resulting constraints---without prior relaxation---are guaranteed to be behaviorally equivalent to the original process model. Especially the presence of \emph{silent transitions}, which are common in real-world process models poses a challenge. Our current approach does not fully address the issues posed by these transitions, with the exception of accounting for optional activities (skip steps) during the translation of relations into constraints.
Further, we acknowledge the limited scope of our relaxation operations, which can be extended to a more comprehensive set.
The validation of our approach was conducted on a single dataset and a more comprehensive evaluation is needed to generalize the findings. Furthermore, the approach's value has not been validated with users, which is an important area for future research.

Future research shall address the limitations outlined above.
Further potential directions for future work include the expansion of relaxation operations, the development of a visual interface to guide domain experts through the relaxation process, and the extension of the translation mechanism to encompass a broader range of process models beyond free-choice workflow nets. 
%
%
%
\bibliographystyle{splncs04}
\bibliography{refs}
\end{document}